\documentclass{iopconfser}

\usepackage{graphicx}
\usepackage{dcolumn}
\usepackage{amssymb,amsmath,bm}
\usepackage{color}
\usepackage[colorlinks,linkcolor=red,citecolor=blue,urlcolor=blue ]{hyperref}
\usepackage{multirow}
\usepackage{enumitem}
\usepackage{mathptmx} 
\usepackage[utf8]{inputenc}
\usepackage{tensor}
\usepackage{amssymb} 
\usepackage{lettrine}
\usepackage[normalem]{ulem}
\usepackage{empheq}
\usepackage{comment}

\begin{document}

\title{Nonlinearities in Gravity: Gravitational Wave Ringdown}

\author{Macarena Lagos$^{1}$}

\affil{$^1$Institute of Astrophysics, Department of Physics and Astronomy, Universidad Andr\'es Bello, Santiago, Chile}

\email{macarena.lagos.u@unab.cl}

\begin{abstract}
The modeling of gravitational wave ringdown has traditionally relied on linear perturbation theory, which mainly describes the late-time behavior of a perturbed black hole after a binary merger. However, the need for more accurate ringdown models has motivated the understanding of nonlinear gravitational effects. In this paper, we summarize the main properties and latest developments of quadratic effects in ringdown models, which are expected to be detectable with next-generation gravitational wave detectors, and will allow for new consistency tests of general relativity.
\end{abstract}

\section{Introduction}
The emission of gravitational waves (GWs) from a remnant black hole (BH), formed after the merger of two compact objects, has been identified for decades as a clean way to perform tests of General Relativity (GR) \cite{1980ApJ...239..292D, Dreyer:2003bv, Cardoso:2016ryw}. This is because perturbed black holes emit GWs at characteristic frequencies, known as quasi-normal modes (QNMs), that only depend on the mass and spin of the BH, according to GR \cite{PhysRevD.2.2141,PhysRev.108.1063, Teukolsky:1973ha, Chandrasekhar:1975zza}. Such a GW signal is known as the ringdown emission, and by measuring some of these ringdown frequencies, one could straightforwardly perform  consistency tests of GR and test for modified gravity (see reviews in \cite{Kokkotas:1999bd, Ferrari:2007dd, Berti:2009kk, Konoplya:2011qq}).

Formally, the ringdown emission is predicted from GR using \textit{linear} perturbation theory around a final stationary remnant BH. At intermediate times\footnote{At early times (i.e.\ soon after the merger) the GW signal is dominated by a prompt response, while at late times the signal is dominated by polynomial tails \cite{PhysRevD.34.384, Andersson:1996cm, PhysRevLett.74.2414}.}, the corresponding GW signal is well approximated by a linear superposition of QNM frequencies $\omega_{\ell m n}$ with amplitudes $A_{\ell m n}$ at the peak of the waveform $u_{\rm peak}$. In geometric units ($c=G=1$), the standard linear ringdown model containing both GW polarizations is the following:
\begin{equation}
    h_+^{(1)}(u,r,\iota,\beta) - ih_\times^{(1)}(u,r,\iota,\beta) = \frac{M}{r}\sum_{\ell m n} A_{\ell m n}\; e^{-i\omega_{\ell m n }(u-u_{\rm peak})}\; {}_{-2}S_{\ell m n}(\iota,\beta; \chi\omega_{\ell m n}), \label{eq:qnms}
\end{equation}
where $u$ is the time variable, $r$ is the distance to the observer, $h_{+,\times}^{(1)}$ are the two linear (and real) GW polarizations in time domain, $M$ is the mass of the remnant BH, and ${}_{-2}S_{\ell m n}(\iota,\beta; \chi\omega_{\ell m n})$ are the spin -2 spheroidal angular harmonics, which are functions of two sky angles $\iota$ and $\beta$, as well as the BH dimensionless spin $\chi$ and the QNM frequencies. 

In Eq.\ (\ref{eq:qnms}), there is a sum over all possible QNMs, which correspond to an infinite discrete spectrum of complex frequencies, $\omega=\omega_R+i\omega_I$, where $\omega_R$ determines the oscillatory frequency of each QNM while $\omega_I$ determines the exponential damping they suffer. These QNMs are labeled by two angular harmonic numbers $(\ell, m)$,  and one overtone integer number $n$ which orders the QNMs from the slowest decaying one ($n=0$) to faster decaying modes (higher $n$ values). The real part $\omega_R$ determines the oscillation timescale of the modes, whereas the complex part $\omega_I<0$ determines their exponential damping timescale. These QNM frequencies depend only on the mass and spin of the remnant BH, and they have been calculated for different BH parameters and harmonic numbers \cite{Berti:2005ys, Berti:2009kk, Stein:2019mop} to use for spectroscopic tests of gravity.

To date, multiple studies have been performed on current data \cite{Capano:2021etf, Finch:2022ynt,Isi:2022mhy, Cotesta:2022pci, Siegel:2023lxl} and forecasts for ground and space-based GW detectors (e.g.\ \cite{Berti:2005ys, Ota:2019bzl, Bhagwat:2019dtm, Pitte:2024zbi}) using linear ringdown QNM frequencies. However, GR is a nonlinear theory and recent studies on second-order perturbation theory  have confirmed the presence of quadratic quasinormal modes (QQNM) in binary black hole (BBH) relativistic numerical simulations \cite{Mitman:2022qdl, Cheung:2022rbm, Ma:2022wpv,Redondo-Yuste:2023seq,Cheung:2023vki, Zhu:2024rej}, extending thus Eq.\ (\ref{eq:qnms}). Next, we will summarize the properties and current status of QQNMs, which is also discussed in a recent ringdown review \cite{Berti:2025hly}.

\section{Quadratic quasi-normal modes}
Black hole perturbation theory up to second order
has the following schematic form:
\begin{eqnarray}
\label{schematic}
\delta_1 G_{\mu\nu} [h^{(1)} ] = 0 \quad ; \quad \delta_1 G_{\mu\nu} [h^{(2)} ]  = - \delta_2 G_{\mu\nu} [h^{(1)}, h^{(1)}] \, ,
\end{eqnarray}
where $h^{(1)}$ and  $h^{(2)}$ denote the first-order and second-order metric perturbation (indices suppressed) around the background black hole spacetime, respectively. In addition, $\delta_1 G_{\mu\nu} $ are the first-order perturbed Einstein equations, representing a linear differential operator which
contains up to two derivatives. The left equation in (\ref{schematic}) leads to the standard Regge-Wheeler \cite{PhysRev.108.1063} and Zerilli \cite{PhysRevD.2.2141} equations for Schwarzschild BHs, or the Teukolsky equation \cite{Teukolsky:1973ha} for Kerr BHs.
The right equation in (\ref{schematic}) will determine the  physics of quadratic QNMs, where one observes that the left-hand side is the same linear operator as that of the linear QNM equations, with the only difference that this is now a inhomogeneous equation. Indeed, in the right-hand side one obtains a source which is bilinear in the first-order perturbations $h^{(1)}$, which comes from perturbing the Einstein equations at second-order, $ \delta_2 G_{\mu\nu}$. The detailed form of this source has been worked out in pioneering papers in \cite{Gleiser:1995gx, Gleiser:1998rw, Brizuela:2009qd}. 

As shown by Eq.\ (\ref{schematic}), the particular solution to $h^{(2)}$ is expected to be fully determined by the quadratic source, and hence by the interactions of two linear, parent, QNMs\footnote{The homogeneous solution to  $h^{(2)}$ is typically ignored as, by construction, has the same form as $h^{(1)}$ and thus can be interpreted as a simple renormalization of the linear QNM amplitudes.}. As a result, the quadratic solution will inherit the properties of the linear QNM solution and take a similar form for asymptotic observers:
\begin{equation}\label{eq:quadratic}
    h_+^{(2)}-ih_\times^{(2)} = \frac{M}{r}\sum_{L M N} A_{L M N }^{(2)}\; e^{-i\omega_{L M N}^{(2)}(u-u_{\rm peak})}\; {}_{-2}S_{L M N}(\iota,\beta),
\end{equation}
where a discrete set of infinite quadratic QNM frequencies is also present, here labeled by $(L M N)$. Importantly, both $A_{L M N }^{(2)}$ and $\omega_{L M N}^{(2)}$ can be determined from the parent linear QNMs. Regarding QQNM frequencies, they are given by \cite{Gleiser:1998rw, Gleiser:1995gx, Ioka:2007ak, Nakano:2007cj,Lagos:2022otp}
\begin{eqnarray}\label{QQNM_freq}
   \omega_{L M N}^{(2)}=  \omega_{\ell m n} + \omega_{\ell' m' n'},
\end{eqnarray}
where the linear parent modes are labeled by $(\ell m n)$ and $(\ell ' m' n')$.
%
%
The exact relationship between the parent numbers $(\ell m n)$ and $(\ell ' m' n')$ to the QQNM numbers $(LMN)$ will be determined by how the two parent spheroidal harmonics source the QQNM spheroidal harmonic. This will be mostly determined by selection rules of angular harmonic multiplications, i.e.
 $M=m+m'$ and $|\ell-\ell'|<L<|\ell+\ell'|$, but a general explicit expression for rotating black holes is not known to date. 

For the most common, nearly-equal mass quasi-circular binaries, the dominant linear QNMs are labeled by $(\ell=2, |m|=2, n=0)$ which are then expected to source the dominant QQNM with a frequency
\begin{eqnarray}\label{Qfreq}
    \omega_{44Q}^{(2)}=  2\omega_{220}.
\end{eqnarray}
From now on this frequency is labeled as $ \omega_{44Q}^{(2)}$ with the subscript $(44Q)$ because it is expected to appear with a largest amplitude in the $(L,|M|)=(4,4)$ angular harmonic of the quadratic solution. The quadratic overtone label $N$ in Eqs.\ (\ref{eq:quadratic})-(\ref{QQNM_freq}) is now simply called $Q$ since an ordering of slowest to fastest decaying QQNMs (analogous to the linear overtone number $n$) is not known yet, so it might be misleading to associate a specific number $N$ to this QQNM frequency.

Regarding QQNM amplitudes, several recent works attempted to obtain the quadratic-to-linear relationship using numerical and analytical methods, obtaining different, even conflicting, results \cite{Kehagias:2023ctr, Redondo-Yuste:2023seq,Cheung:2023vki, Zhu:2024rej, Ma:2024qcv}. However, recent studies \cite{Bourg:2024jme, Bucciotti:2024jrv} have shown that it depends on the amplitude of parity odd and even linear QNMs, with \cite{Bourg:2024jme, Khera:2024yrk} showing that the previous conflicting results were due to different QNM parity choices. 

When the selected parent modes are $(\ell=2, |m|=2, n=0)$ and the binary is not precessing (and it is thus fully parity even due to planar symmetry), the sourced QQNM with frequency (\ref{Qfreq}) will have an amplitude \cite{Bucciotti:2024zyp,Khera:2024yrk, Bucciotti:2024jrv} 
\begin{eqnarray}\label{Qamp}
    A^{(2)}_{44Q}=0.154 e^{-i0.068 } \; (A_{220})^2.
\end{eqnarray}
Note that Eq.\ (\ref{Qamp}) was calculated for a final non-spinning BH, yet most remnant BHs from mergers are expected to have final dimensionless spins over $\chi>0.6$. While the 
quadratic form of Eq.\ (\ref{Qamp}) is always valid, the constant factors relating $ A^{(2)}_{44Q}$ and $A_{220}^2$ will depend on spin, albeit weakly \cite{Khera:2024yrk}. 

Importantly, the fact that QQNMs have both their frequency and amplitude fully determined by their parent linear modes will mean that QQNMs can be used to test GR in two ways: by measuring both their frequency and testing the relationship in Eq.\ (\ref{Qfreq}); or by measuring their amplitude and testing Eq.\ (\ref{Qamp}).

\section{Quadratic QNMs in binary mergers}
In order to confirm the presence of QQNMs in binary merger signals, one can analyze numerical relativity simulations which solve the exact Einstein equations and hence contain all nonlinear effects.

In the seminal papers \cite{Mitman:2022qdl, Cheung:2022rbm} different numerical simulations of black hole binary mergers from the SXS catalog \cite{Boyle:2019kee} were fitted with a ringdown model with QQNMs. For concreteness, let us consider the case of the SXS simulation with ID BBHX:0305, which mimics the first direct GW event detected GW150914 \cite{LIGOScientific:2016vbw}. In this case, the GW signal $h_+-ih_\times$ after the peak is extracted, and its contribution to the $(4,4)$ angular harmonic is fitted by two ringdown models:
\begin{align}
    h_{(4,4)}^{\rm model, L}(u)= \frac{M}{r}\sum_{n=0}^{2}A_{44n}\,e^{-i\omega_{44n}(u-u_{\rm peak})}; \; h_{(4,4)}^{\rm model, Q}(u)= \frac{M}{r}\left(\sum_{n=0}^{1}A_{44n}\,e^{-i\omega_{44n}(u-u_{\rm peak})} + A^{(2)}_{44Q}\,e^{-i\omega^{(2)}_{44Q}(u-u_{\rm peak})}\right).
\end{align}
Here, the model $h_{(4,4)}^{\rm model, L}$ contains three QNMs with overtone numbers $n=0,1,2$ and it is purely linear since the frequencies $\omega_{44n}$ will be fixed by their linear QNM values for the best-fit remnant mass and spin of GW150914. On the other hand, the model $h_{(4,4)}^{\rm model, Q}$ also contains a total of three QNMs but two are linear modes with overtones $n=0,1$ and the third mode is a purely QQNM corresponding to the $(44Q)$ mode previously discussed, with fixed frequency $\omega^{(2)}_{44Q}=2\omega_{220}$. In both models, the amplitudes $A_{44n}$ and $A^{(2)}_{44Q}$ will be fitted by a least-square minimization procedure. 
\begin{figure}[h!]
\centering
\includegraphics[width=0.50\linewidth]{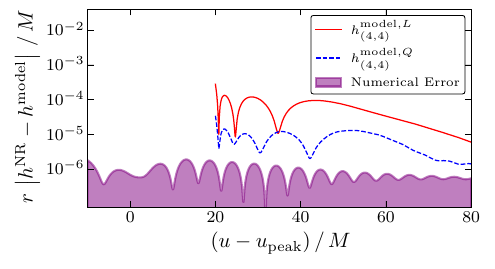}
\caption{\label{fig:residual} Residual of the simulation for a GW150914-like event and two ringdown models: a purely linear model (red) and a model with QQNM (blue). Fig.\ from \cite{Mitman:2022qdl}.}
\end{figure}
The residuals of both models are shown in Fig.\ \ref{fig:residual}, when fitting the simulation at a time $u-u_{\rm peak}=20M$. Crucially, the residual of the linear model is one order of magnitude larger than that of the quadratic model, confirming thus the importance of the QQNM. Other tests of mismatch, frequency fitting, and amplitude scaling are performed in  \cite{Mitman:2022qdl}, which confirm the robust presence of this QQNM in the simulation. In this event, the QQNM amplitude is estimated to be about $A^{(2)}_{44Q}\approx 10\% A_{220}$, showing that this QQNM is indeed a perturbative effect yet definitely \textit{not negligible}. 

Fig.\ \ref{fig:residual} shows the residuals $20M$ after the merger because at earlier times \cite{Mitman:2022qdl} shows that the mismatch is large and it seems like both of these ringdown models only fit well the simulation from around $20M$ onward. As a result, the inclusion of second-order QNM effects mostly improve ringdown models are \textit{late times}, not early times.

While here we have shown the results for only one binary simulation, \cite{Mitman:2022qdl, Cheung:2022rbm} analyzed various other simulations and obtained qualitatively similar results. 
Other analyses of numerical binary simulations have confirmed the presence of other QQNMs \cite{Ma:2022wpv, Cheung:2023vki, Giesler:2024hcr, Mitman:2025hgy}, including quadratic effects present in the $(2,2)$ angular harmonic which could make percent-level contributions to the total ringdown signal.

\section{Detectability of Quadratic QNMs}
The dominant QQNM expected in non-precessing nearly equal-mass binaries mentioned above has been shown to make $10\%$ contributions to the signal \cite{Mitman:2022qdl, Cheung:2022rbm}, and it is thus likely detectable in the future. If detected, QQNMs would allow to test the nonlinear behavior of GR around black holes. 

For concreteness, let us consider again an event similar to GW150914, with $z=0.093$, final remnant mass $M=61M_\odot$ and spin $\chi=0.69$. By fitting a ringdown model to the after-merger signal from the BBH:0305 simulation, we can estimate the signal-to-noise ratio (SNR) expected in future GW detectors. 
\begin{figure}[h!]
\centering
\includegraphics[width=0.45\linewidth]{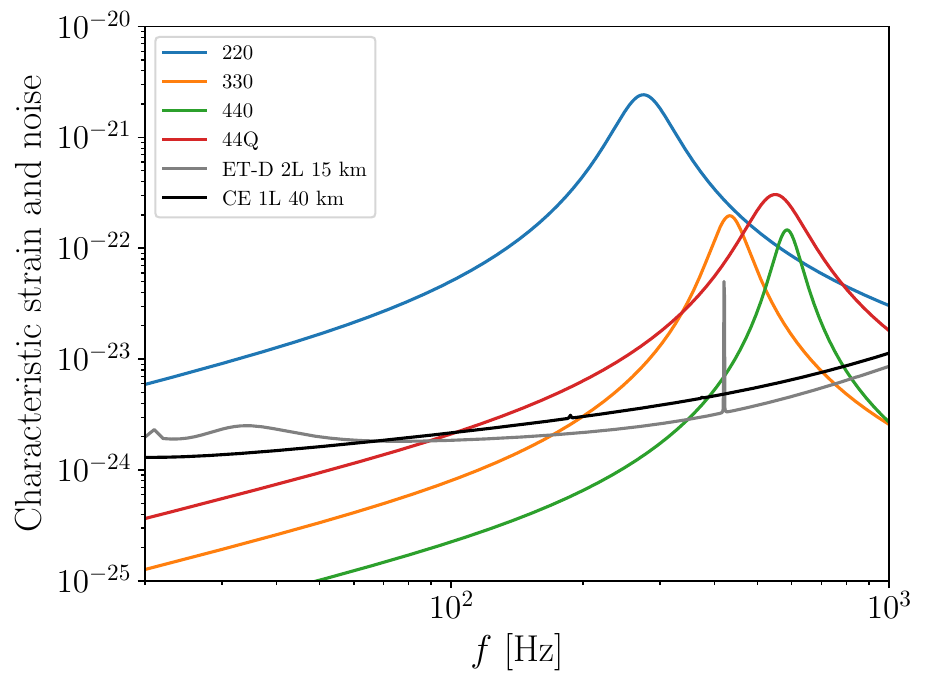}
\caption{\label{fig:ET}  Characteristic noise (black and gray lines) for ET and CE detectors, compared to characteristic strain (colored lines) of the individual QNMs for a GW150914-like GW event. Fig.\ from \cite{Lagos:2024ekd}.}
\end{figure}
Fig.\ \ref{fig:ET} shows the characteristic noise for Einstein Telescope (ET) \cite{Branchesi:2023mws} and Cosmic Explorer (CE) \cite{Evans:2021gyd}, compared to the characteristic strain of individual QNMs expected to be important in future detections: namely the linear modes $(220), (330), (440)$ and the QQNM $(44Q)$. Here it is visually clear that the QQNM (in red) will have a larger SNR than the linear modes  $(330), (440)$. Indeed, for this event at redshift $z=0.093$ we obtain the following SNR for each QNM: $\rho_{220}=199$, $\rho_{330}=11.3$, $\rho_{440}=5.68$, and $\rho_{44Q}=18.1$  for ET (and comparable values for CE). We thus confirm that the QQNM has an SNR \textit{larger} than the (330) and (440) linear modes, making it more easily detectable. 

The work in \cite{Yi:2024elj} performed a binary population study and obtained that CE and ET detectors could observe this QQNM in up to a few tens of events per year with SNR $\rho_{44Q}>8$. On the other hand, the detectability of this QQNM in the future space-based detector LISA \cite{Colpi:2024xhw} was studied in \cite{Yi:2024elj, Shi:2024ttu}. A population study showed that LISA could detect this QQNM with $\rho_{44Q}>8$ in up to thousands of events during a four-year mission. Furthermore, based again on this SNR threshold, a recent study analyzing the detectability of subdominant quadratic QNMs \cite{Khera:2024yrk} claims that several other QQNMs could be observed by CE and LISA.

Next, we can quantify exactly how well the QQNM parameters (frequency and amplitude) will be constrained with future GWs, by performing a fisher forecast. As discussed in \cite{Lagos:2024ekd}, imposing a QQNM threshold of 8 is simply a necessary condition since the detectability of multiple QNMs has further requirements. Indeed, for a GW event like GW150914 we find that the $(440)$ and $(44Q)$ modes are highly correlated, which causes considerable degradation in the future measurement of their frequencies. 
If one wishes to distinguish the two modes $(440)$ and $(44Q)$ and confirm their amplitude to be non-zero at 68\%CL, one needs to increase their SNR even more. 

\begin{figure}[h!]
\centering
\includegraphics[width=0.45\linewidth]{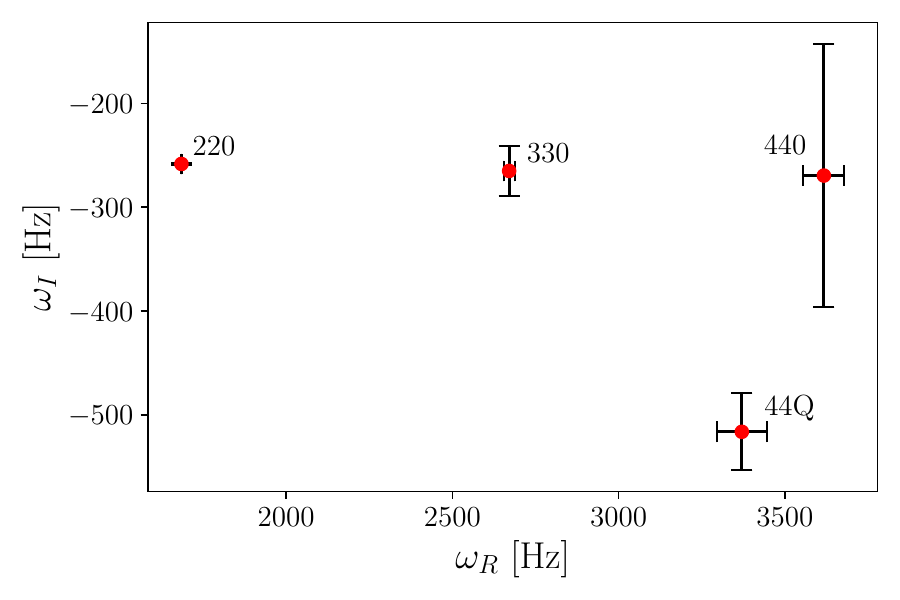}
\includegraphics[width=0.45\linewidth]{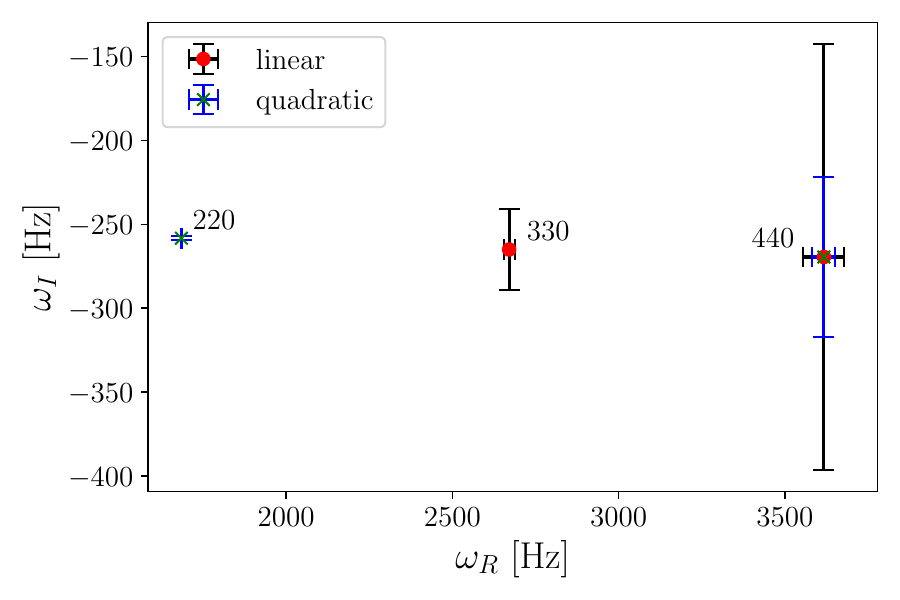}
\caption{\label{fig:0305} Real and imaginary components of the QNM frequencies (in Hz) and their $1\sigma$ uncertainty, for a GW150914-like event at $z=0.045$ in ET. Left shows the errors when the QQNM is measured independently. Right shows the errors when the QQNM is used as a dependent mode (blue) and compares it to the independent case (black). Figs.\ from \cite{Lagos:2024ekd}.}
\end{figure}

Fig.\ \ref{fig:0305} shows on the left the precision expected on the real and imaginary frequencies of four QNMs, for an event like GW150914 but placed at redshift $z=0.045$ instead of $z=0.093$ to increase its SNR. Here we see that the QQNM frequency is measured worse than that of the (330) mode, but better than the (440) mode. This is expected to be a general trend for non-precessing quasi-circular nearly equal-mass binary mergers.

While these previous results show that the QQNM can be measured independently in order to perform future tests of gravity, they can also be used in a different way to perform consistency test of GR. Using the predictions of GR in Eqs.\ (\ref{Qfreq}) and (\ref{Qamp}) it is possible to construct a ringdown model where the QQNM is used as a dependent mode:
\begin{equation}
    h_{(4,4)}^{\rm model, Q}(u)= \frac{M}{r}\left(\sum_{n=0}^{1}A_{44n}\,e^{-i\omega_{44n}(u-u_{\rm peak})} +0.154\,(A_{220})^{2}\,e^{-i0.068}e^{-2i\omega_{220}(u-u_{\rm peak})}\right)
\end{equation}
in order to \textit{improve measurements on its parent modes and to reduce parameter degeneracy} with other QNMs. Fig.\ \ref{fig:0305} shows on the right the improved constraints (blue) in the (220) and (440) modes when using this new ringdown model. The presence of the QQNM expressed in a dependent way is shown to improve the (220) frequency parameters by a small factor of about $\times 1.05$ compared to the independent QQNM model. This improvement is small because the (220) mode has very high SNR on its own. Surprisingly, the dependent QQNM mostly helps break degeneracies with the parameters of the (440) mode and thus help find better constraints in the (440) frequency by a factor of $\times 2$. The presence of the QQNM in the model will then help perform more precise consistency tests of GR with linear modes.

Finally, we notice that so far the examples shown here concern events akin to GW150914, but for unequal-mass binaries this QQNM is expected to have a smaller effect \cite{Mitman:2022qdl}. The study in \cite{Lagos:2024ekd} also considered the case of intermediate-mass black holes (IMBHs) whose higher mass makes the ringdown signal peak at a frequency range where ET and CE have lower noise. Moreover, while the evolutionary formation history of IMBHs is not known, it is expected that some of them will have larger spins, which has been shown to increase the relative importance of the QQNM \cite{Cheung:2023vki, Yi:2024elj}. If one considers an event that mimics  GW190521 \cite{LIGOScientific:2020iuh}, the fisher forecast analysis shows that some QQNM parameters can be measured even better than those of the (330) mode. Therefore, it is crucial to consider this QQNM when analyzing the ringdown of IMBH systems, such as GW190521 or the recently confirmed event GW231123 \cite{LIGOScientific:2025rsn}.

\section{Conclusions}
While black hole perturbation theory works well for modeling the ringdown at late times, second order effects are not negligible. In the past few years, several works have shown the importance of a specific quadratic QNM relevant for nearly equal-mass quasi-circular non-precessing binary systems. This QQNM is found to be as large as $10\%$ of the total signal and will appear in the angular harmonic $(\ell, |m|)=(4,4)$. Due to its large effect, it will be detectable by next-generation ground based and space-based detectors. Interestingly, QQNMs can be used to perform regular spectroscopic tests of gravity, by measuring their frequency independently, but they can also be used as a way to improve the precision on the measurement of linear QNMs parameters and hence to perform more precise tests of GR. This can be done by using the GR predictions on both frequency and amplitude of QQNMs.

The inclusion of nonlinear effects in ringdown is still not fully explored and open questions remain. On the one hand, forecast studies on other QQNMs (especially those present in the $(\ell,|m|)=(2,2)$ angular harmonic) remain to be performed. In addition, it is still not known what modified gravity theories predict for these QQNMs and hence what to expect in future spectroscopic tests of gravity. On the other hand, other types of nonlinearities might be important to include in ringdown models as well, such as transient changes in mass and spin after the merger \cite{Sberna:2021eui}, since those might potentially improve the ringdown model close to the merger and hence exploit high SNR regions of the data.

\section*{Acknowledgements}
M.L.'s participation in the GR 24 \& Amaldi 16 conference was partially supported by the Institute of Physics (IOP) and Fondecyt Iniciaci\'on Chile grant 11250105.

\bibliography{References}

\providecommand{\newblock}{}
\begin{thebibliography}{10}
\expandafter\ifx\csname url\endcsname\relax
  \def\url#1{{\tt #1}}\fi
\expandafter\ifx\csname urlprefix\endcsname\relax\def\urlprefix{URL }\fi
\providecommand{\eprint}[2][]{\url{#2}}

\bibitem{1980ApJ...239..292D}
{Detweiler} S 1980 {\em The Astrophysical Journal\/} {\bf 239} 292--295

\bibitem{Dreyer:2003bv}
Dreyer O, Kelly B~J, Krishnan B, Finn L~S, Garrison D and Lopez-Aleman R 2004 {\em Class. Quant. Grav.\/} {\bf 21} 787--804 (\textit{Preprint} \eprint{gr-qc/0309007})

\bibitem{Cardoso:2016ryw}
Cardoso V and Gualtieri L 2016 {\em Class. Quant. Grav.\/} {\bf 33} 174001 (\textit{Preprint} \eprint{1607.03133})

\bibitem{PhysRevD.2.2141}
Zerilli F~J 1970 {\em Phys. Rev. D\/} {\bf 2}(10) 2141--2160 \urlprefix\url{https://link.aps.org/doi/10.1103/PhysRevD.2.2141}

\bibitem{PhysRev.108.1063}
Regge T and Wheeler J~A 1957 {\em Phys. Rev.\/} {\bf 108}(4) 1063--1069 \urlprefix\url{https://link.aps.org/doi/10.1103/PhysRev.108.1063}

\bibitem{Teukolsky:1973ha}
Teukolsky S~A 1973 {\em Astrophys. J.\/} {\bf 185} 635--647

\bibitem{Chandrasekhar:1975zza}
Chandrasekhar S and Detweiler S~L 1975 {\em Proc. Roy. Soc. Lond. A\/} {\bf 344} 441--452

\bibitem{Kokkotas:1999bd}
Kokkotas K~D and Schmidt B~G 1999 {\em Living Rev. Rel.\/} {\bf 2} 2 (\textit{Preprint} \eprint{gr-qc/9909058})

\bibitem{Ferrari:2007dd}
Ferrari V and Gualtieri L 2008 {\em Gen. Rel. Grav.\/} {\bf 40} 945--970 (\textit{Preprint} \eprint{0709.0657})

\bibitem{Berti:2009kk}
Berti E, Cardoso V and Starinets A~O 2009 {\em Class. Quant. Grav.\/} {\bf 26} 163001 (\textit{Preprint} \eprint{0905.2975})

\bibitem{Konoplya:2011qq}
Konoplya R~A and Zhidenko A 2011 {\em Rev. Mod. Phys.\/} {\bf 83} 793--836 (\textit{Preprint} \eprint{1102.4014})

\bibitem{PhysRevD.34.384}
Leaver E~W 1986 {\em Phys. Rev. D\/} {\bf 34}(2) 384--408 \urlprefix\url{https://link.aps.org/doi/10.1103/PhysRevD.34.384}

\bibitem{Andersson:1996cm}
Andersson N 1997 {\em Phys. Rev. D\/} {\bf 55} 468--479 (\textit{Preprint} \eprint{gr-qc/9607064})

\bibitem{PhysRevLett.74.2414}
Ching E~S~C, Leung P~T, Suen W~M and Young K 1995 {\em Phys. Rev. Lett.\/} {\bf 74}(13) 2414--2417 \urlprefix\url{https://link.aps.org/doi/10.1103/PhysRevLett.74.2414}

\bibitem{Berti:2005ys}
Berti E, Cardoso V and Will C~M 2006 {\em Phys. Rev. D\/} {\bf 73} 064030 (\textit{Preprint} \eprint{gr-qc/0512160})

\bibitem{Stein:2019mop}
Stein L~C 2019 {\em J. Open Source Softw.\/} {\bf 4} 1683 (\textit{Preprint} \eprint{1908.10377})

\bibitem{Capano:2021etf}
Capano C~D, Cabero M, Westerweck J, Abedi J, Kastha S, Nitz A~H, Wang Y~F, Nielsen A~B and Krishnan B 2023 {\em Phys. Rev. Lett.\/} {\bf 131} 221402 (\textit{Preprint} \eprint{2105.05238})

\bibitem{Finch:2022ynt}
Finch E and Moore C~J 2022 {\em Phys. Rev. D\/} {\bf 106} 043005 (\textit{Preprint} \eprint{2205.07809})

\bibitem{Isi:2022mhy}
Isi M and Farr W~M 2023 {\em Phys. Rev. Lett.\/} {\bf 131}(16) 169001 (\textit{Preprint} \eprint{2202.02941}) \urlprefix\url{https://link.aps.org/doi/10.1103/PhysRevLett.131.169001}

\bibitem{Cotesta:2022pci}
Cotesta R, Carullo G, Berti E and Cardoso V 2022 {\em Phys. Rev. Lett.\/} {\bf 129} 111102 (\textit{Preprint} \eprint{2201.00822})

\bibitem{Siegel:2023lxl}
Siegel H, Isi M and Farr W~M 2023 {\em Phys. Rev. D\/} {\bf 108} 064008 (\textit{Preprint} \eprint{2307.11975})

\bibitem{Ota:2019bzl}
Ota I and Chirenti C 2020 {\em Phys. Rev. D\/} {\bf 101} 104005 (\textit{Preprint} \eprint{1911.00440})

\bibitem{Bhagwat:2019dtm}
Bhagwat S, Forteza X~J, Pani P and Ferrari V 2020 {\em Phys. Rev. D\/} {\bf 101} 044033 (\textit{Preprint} \eprint{1910.08708})

\bibitem{Pitte:2024zbi}
Pitte C, Baghi Q, Besan{\c{c}}on M and Petiteau A 2024 {\em Phys. Rev. D\/} {\bf 110} 104003 (\textit{Preprint} \eprint{2406.14552})

\bibitem{Mitman:2022qdl}
Mitman K {\em et~al.\/} 2023 {\em Phys. Rev. Lett.\/} {\bf 130} 081402 (\textit{Preprint} \eprint{2208.07380})

\bibitem{Cheung:2022rbm}
Cheung M~H~Y {\em et~al.\/} 2023 {\em Phys. Rev. Lett.\/} {\bf 130} 081401 (\textit{Preprint} \eprint{2208.07374})

\bibitem{Ma:2022wpv}
Ma S, Mitman K, Sun L, Deppe N, H\'ebert F, Kidder L~E, Moxon J, Throwe W, Vu N~L and Chen Y 2022 {\em Phys. Rev. D\/} {\bf 106} 084036 (\textit{Preprint} \eprint{2207.10870})

\bibitem{Redondo-Yuste:2023seq}
Redondo-Yuste J, Carullo G, Ripley J~L, Berti E and Cardoso V 2024 {\em Phys. Rev. D\/} {\bf 109} L101503 (\textit{Preprint} \eprint{2308.14796})

\bibitem{Cheung:2023vki}
Cheung M~H~Y, Berti E, Baibhav V and Cotesta R 2024 {\em Phys. Rev. D\/} {\bf 109} 044069 (\textit{Preprint} \eprint{2310.04489})

\bibitem{Zhu:2024rej}
Zhu H {\em et~al.\/} 2024 {\em Phys. Rev. D\/} {\bf 109} 104050 (\textit{Preprint} \eprint{2401.00805})

\bibitem{Berti:2025hly}
Berti E {\em et~al.\/} 2025  (\textit{Preprint} \eprint{2505.23895})

\bibitem{Gleiser:1995gx}
Gleiser R~J, Nicasio C~O, Price R~H and Pullin J 1996 {\em Class. Quant. Grav.\/} {\bf 13} L117--L124 (\textit{Preprint} \eprint{gr-qc/9510049})

\bibitem{Gleiser:1998rw}
Gleiser R~J, Nicasio C~O, Price R~H and Pullin J 2000 {\em Phys. Rept.\/} {\bf 325} 41--81 (\textit{Preprint} \eprint{gr-qc/9807077})

\bibitem{Brizuela:2009qd}
Brizuela D, Martin-Garcia J~M and Tiglio M 2009 {\em Phys. Rev. D\/} {\bf 80} 024021 (\textit{Preprint} \eprint{0903.1134})

\bibitem{Ioka:2007ak}
Ioka K and Nakano H 2007 {\em Phys. Rev. D\/} {\bf 76} 061503 (\textit{Preprint} \eprint{0704.3467})

\bibitem{Nakano:2007cj}
Nakano H and Ioka K 2007 {\em Phys. Rev. D\/} {\bf 76} 084007 (\textit{Preprint} \eprint{0708.0450})

\bibitem{Lagos:2022otp}
Lagos M and Hui L 2023 {\em Phys. Rev. D\/} {\bf 107} 044040 (\textit{Preprint} \eprint{2208.07379})

\bibitem{Kehagias:2023ctr}
Kehagias A, Perrone D, Riotto A and Riva F 2023 {\em Phys. Rev. D\/} {\bf 108} L021501 (\textit{Preprint} \eprint{2301.09345})

\bibitem{Ma:2024qcv}
Ma S and Yang H 2024 {\em Phys. Rev. D\/} {\bf 109} 104070 (\textit{Preprint} \eprint{2401.15516})

\bibitem{Bourg:2024jme}
Bourg P, Panosso~Macedo R, Spiers A, Leather B, Bonga B and Pound A 2025 {\em Phys. Rev. Lett.\/} {\bf 134} 061401 (\textit{Preprint} \eprint{2405.10270})

\bibitem{Bucciotti:2024jrv}
Bucciotti B, Juliano L, Kuntz A and Trincherini E 2024 {\em JHEP\/} {\bf 09} 119 (\textit{Preprint} \eprint{2406.14611})

\bibitem{Khera:2024yrk}
Khera N, Ma S and Yang H 2025 {\em Phys. Rev. Lett.\/} {\bf 134} 211404 (\textit{Preprint} \eprint{2410.14529})

\bibitem{Bucciotti:2024zyp}
Bucciotti B, Juliano L, Kuntz A and Trincherini E 2024 {\em Phys. Rev. D\/} {\bf 110} 104048 (\textit{Preprint} \eprint{2405.06012})

\bibitem{Boyle:2019kee}
Boyle M {\em et~al.\/} 2019 {\em Class. Quant. Grav.\/} {\bf 36} 195006 (\textit{Preprint} \eprint{1904.04831})

\bibitem{LIGOScientific:2016vbw}
Abbott B~P {\em et~al.\/} (LIGO Scientific, Virgo) 2016 {\em Phys. Rev. D\/} {\bf 93} 122003 (\textit{Preprint} \eprint{1602.03839})

\bibitem{Giesler:2024hcr}
Giesler M {\em et~al.\/} 2025 {\em Phys. Rev. D\/} {\bf 111} 084041 (\textit{Preprint} \eprint{2411.11269})

\bibitem{Mitman:2025hgy}
Mitman K {\em et~al.\/} 2025 {\em Phys. Rev. D\/} {\bf 112} 064016 (\textit{Preprint} \eprint{2503.09678})

\bibitem{Lagos:2024ekd}
Lagos M, Andrade T, Rafecas-Ventosa J and Hui L 2025 {\em Phys. Rev. D\/} {\bf 111} 024018 (\textit{Preprint} \eprint{2411.02264})

\bibitem{Branchesi:2023mws}
Branchesi M {\em et~al.\/} 2023 {\em JCAP\/} {\bf 07} 068 (\textit{Preprint} \eprint{2303.15923})

\bibitem{Evans:2021gyd}
Evans M {\em et~al.\/} 2021  (\textit{Preprint} \eprint{2109.09882})

\bibitem{Yi:2024elj}
Yi S, Kuntz A, Barausse E, Berti E, Cheung M~H~Y, Kritos K and Maselli A 2024 {\em Phys. Rev. D\/} {\bf 109} 124029 (\textit{Preprint} \eprint{2403.09767})

\bibitem{Colpi:2024xhw}
Colpi M {\em et~al.\/} 2024  (\textit{Preprint} \eprint{2402.07571})

\bibitem{Shi:2024ttu}
Shi C, Zhang Q and Mei J 2024 {\em Phys. Rev. D\/} {\bf 110} 124007 (\textit{Preprint} \eprint{2407.13110})

\bibitem{LIGOScientific:2020iuh}
Abbott R {\em et~al.\/} (LIGO Scientific, Virgo) 2020 {\em Phys. Rev. Lett.\/} {\bf 125} 101102 (\textit{Preprint} \eprint{2009.01075})

\bibitem{LIGOScientific:2025rsn}
Collaboration T~L~S, the Virgo~Collaboration and the KAGRA~Collaboration (LIGO Scientific, VIRGO, KAGRA) 2025  (\textit{Preprint} \eprint{2507.08219})

\bibitem{Sberna:2021eui}
Sberna L, Bosch P, East W~E, Green S~R and Lehner L 2022 {\em Phys. Rev. D\/} {\bf 105} 064046 (\textit{Preprint} \eprint{2112.11168})

\end{thebibliography}

\end{document}